# Inner/Outer Ratio Similarity Scaling for 2-D Wall-bounded Turbulent Flows

By David W. Weyburne
*Air Force Research Laboratory, 2241 Avionics Circle, Wright-Patterson AFB, OH 45433, USA*

**Abstract**

The turbulent boundary layer scaling parameters for the velocity profile are usually associated with either the inner viscous region or the outer boundary layer region. It has been a long-held view that complete similarity of the velocity profile can only occur if the inner and outer region scaling parameters change proportionally as one moves from station to station along the wall. However, it appears that complete similarity is not possible for the wall-bounded turbulent boundary layer. Hence, the outer/inner ratio would seem to be of little use. However, recent revelations revive the need for identifying likely experimental datasets that display outer region similarity. It is our contention that likely datasets can be identified by finding datasets in which the inner-outer thickness ratio is almost constant. This inner-outer thickness ratio is usually associated with the Rotta scaling ratio. Unfortunately, the Rotta ratio proportional change condition has never been shown to be a similarity requirement. In contrast, we show that a recently developed thickness ratio based on the integral moment method must change proportionately from station to station if similarity is present.

## 1. INTRODUCTION

One of the most fundament concepts in fluid mechanic's is the search for ways to analyze experiment observables using dimensional analysis with the intent of finding scaling parameters that render the scaled observable from different stations along the flow to appear to be similar. Similarity of the velocity profile formed by fluid flow along a wall is one of those fundamental concepts. For 2-D wall-bounded flows, velocity profile similarity is defined as the case where two velocity profiles taken at different stations along the flow differ only by simple scaling parameters in $y$ and $u(x,y)$, where $y$ is the normal direction to the wall, $x$ is the flow direction, and $u(x,y)$ is the velocity parallel to the wall in the flow direction. The velocity profile is defined as the velocity $u(x,y)$ at a fixed x for all y. Similarity solutions of the flow governing equations are well known for laminar flow. Turbulent flow similarity is problematic. One of the complicating factors for the turbulent boundary layer is that it appears to be composed of two distinct regions, the near wall region where viscosity effects are important and the outer region where they are not.

In the past, it was generally assumed that the complete profile similarity could only occur if the inner/outer thickness ratio is constant from station to station. However, experimental and theoretical evidence [1] seems to indicate that the wall-bounded turbulent boundary layer velocity profiles do not show whole profile similarity [2]. Is it even relevant to consider the thickness ratio anymore? Under the current paradigm, the answer would seem to be no. However, there has been a recent revelation which changes this dynamic. According to Castillo, George, and co-workers (see [3-5] and references therein), most outer regions display similarity when scaled with the Zagarola and Smits [6] parameter set. If this were true then there would be no need to filter experimental datasets for similarity; it would be a given that similarity is probably present. However, recent work by Weyburne [7-9] casts serious doubt on Castillo, George, and co-workers similarity prevalence claim. The problem, as outlined by Weyburne, is that the Zagarola and Smits velocity scaling parameter, $u_{ZS}$, does not always satisfy all of the defect profile similarity requirements. What has been missed in the literature is a realization that every defect profile based theoretical paper on wall-bounded turbulent boundary layer similarity [1,3,10,11] has indicated that similarity requires any candidate scaling parameter, such as $u_{ZS}$, must be proportional to $u_e$, the velocity at the boundary layer edge, at each measurement station. This result holds even when one only considers similarity in the outer region [3]. Unfortunately, for whatever reason, this similarity requirement has been ignored. When Weyburne [7,8] cross-checked many similarity claims from the literature, it was found that this condition was NOT satisfied. Hence, according to Weyburne, most experimental wall-bounded turbulent boundary layer datasets do NOT show similarity in the outer region. This is consistent with earlier experimental efforts [12,13] which noted that creating similarity conditions in the wind tunnel is very difficult.

Therefore, if turbulent boundary layer similarity is relatively rare, then it is once again relevant to ask under what conditions can one expect to see similarity in the outer region of the turbulent boundary layer. One search option would be to follow Clauser [12] and Castillo and George [3] and use a pressure gradient based criteria obtained as a requirement for similarity in the flow governing approach to similarity. While this may in fact be an acceptable approach for that turbulent boundary layers which have pressure gradients in the flow direction, this approach fails for the zero-pressure gradient case. As an alternative, we propose to revive the thickness ratio as a criterion for similarity in the outer region. Our conjecture is that outer region similarity will only be observed when the inner/outer thickness ratio is "almost" constant.

Previously, this outer/inner ratio thickness test has been based on checking the Rotta [10] ratio $u_e/u_\tau$, where $u_e$ is the velocity at the boundary layer edge and $u_\tau$ is the friction velocity. The Rotta ratio assumes that the thicknesses of the outer and inner regions are proportional to the similarity scaling velocities for the two regions. However, this assumption has never been proven. In particular, $u_\tau$ has never been shown theoretically to be a valid similarity scaling parameter for the inner region (see Appendix A). In addition, the assumption that all inner regions of wall-bounded turbulent boundary layers are similar using the Prandtl Plus parameters, $\nu/u_\tau$ and $u_\tau$, has recently been challenged by Weyburne [14]. Using a flow governing approach to similarity, he showed that the only turbulent flow situation that shows similarity with the Prandtl Plus scaling parameters is for turbulent sink flows. The Prandtl Plus parameters are NOT similarity scaling parameters for the more general Falkner-Skan flows along a wall. For these two reasons, the Rotta ratio may not be the appropriate way to determine if the inner and outer region thicknesses are changing proportionately.

There is an alternative thickness ratio test. A relatively new integral moment method for describing the boundary thickness and shape has been developed [15,16]. Borrowing from probability density function (PDF) methodology, the thickness and shape of the velocity profile are described in terms of integral moments of the velocity and its first two derivatives. This method provides an experimentally accessible measurement of the inner boundary layer thickness, $\delta_v$ and an outer boundary layer thickness, $\delta_d$. In this work, we show that this new boundary layer thickness ratio is a similarity requirement: if similarity is present in a set velocity profiles then the ratio $\delta_v/\delta_d$ must be constant from station to station. The same cannot be said of the Rotta constraint.

## 2. Boundary Layer Description

In this section, the new boundary layer description method developed by Weyburne [15,16] is briefly outlined. The new boundary layer description borrows from PDF methodology and is based on integral moments of the stream-wise velocity $u(x,y)$ and its first two derivatives in the direction normal to the wall (y-direction). The relevant boundary layer parameters are described for both the inner region and outer region of the turbulent boundary layer.

### 2.1 Viscous Moments

The inner boundary layer region description is based on central moments of the second derivative kernel. Since the second derivative of the velocity is directly related to the viscous term in the momentum balance equation, we have termed this set of moments the "viscous" boundary layer moments. For wall-bounded 2-D flow over a flat wall, the viscous velocity boundary layer $n$th central moment, $\lambda_n$ is defined as

$$\lambda_n(x) = \int_0^h dy\, (y-\mu_1)^n \frac{d^2\{-\mu_1 u(x,y)/u_e\}}{dy^2} \,, \tag{1}$$

where $y=h$ is deep into the free stream, $u_e(x)$ is the free stream velocity at the boundary layer edge, and where the first moment about the origin, $\mu_1(x)$, is also the normalizing parameter. The derivative in Eq. 1 is written in this way to emphasize the probability-density-function-like behavior. Now it is straightforward to show that

$$\mu_1(x) = \frac{u_e}{\left.\frac{du(x,y)}{dy}\right|_{y=0}} = \frac{\nu u_e}{u_\tau^2} \,, \tag{2}$$

where $\nu$ is the kinematic viscosity. The first moment about the origin, $\mu_1(x)$, is referred to as the viscous mean location. This parameter is a characteristic parameter of any boundary layer region where viscous forces are present, including the inner region of the turbulent boundary layer.

Borrowing from PDF methodology, the inner region thickness $\delta_v$ is given by

$$\delta_v(x) = \mu_1 + 2\sigma_v \,, \tag{3}$$

where the viscous boundary layer width is given by $\sigma_v(x) = \sqrt{\lambda_2} = \sqrt{-\mu_1^2 + 2\mu_1\delta_1}$ and the displacement thickness, $\delta_1(x)$, is given by

$$\delta_1(x) = \int_0^h dy \, \{1 - u(x,y)/u_e\} \, . \tag{4}$$

This means that the inner region parameter values for $\mu_1$, $\sigma_v$, and $\delta_v$ can be calculated in terms of experimentally accessible data. Note that the probability community sometimes uses the mean plus three or four times $\sigma_v$ instead of two-sigma as used herein. The two-sigma value was chosen so that $\delta_v$ approximately tracks the 99% thickness $\delta_{99}$ for laminar flow.

## 2.2 First Derivative Moments

In addition to the second derivative kernel, other integral kernels for describing the boundary layer are possible. The kernel based on the stream-wise velocity itself and another kernel based on the derivative of the stream-wise flow velocity in the direction normal to the plate are obvious choices for describing the outer region of the turbulent boundary layer. For convenience, we will use the first derivative kernel herein. The derivative of the stream-wise velocity boundary layer $n$th moment, $\kappa_n$ is defined as

$$\kappa_n(x) = \int_0^h dy \, (y - \delta_1)^n \, \frac{d\{u(x,y)/u_e\}}{dy} \, , \tag{5}$$

such that $y = h$ is deep into the free stream.

The definition of the stream-wise boundary layer thickness $\delta_d$ is defined as the point at which the stream-wise velocity derivative becomes negligible. Following standard PDF protocol, the thickness is taken as

$$\delta_d(x) = \delta_1 + 4\sigma_s \, , \tag{6}$$

where the stream-wise velocity boundary layer width is defined as $\sigma_s(x) = \sqrt{\kappa_2} = \sqrt{-\delta_1^2 + 2\alpha_1}$ and where

$$\alpha_1(x) = \int_0^h dy \, y\{1 - u(x,y)/u_e\} \, . \tag{7}$$

Notice that the $\alpha_1$ parameter is very similar to the displacement thickness. This means that all of the outer region $\delta_1$, $\sigma_s$, and $\delta_d$ values can be calculated in terms of experimental velocity profile data.

The parameters based on $\kappa_n$ moments describe the thickness and shape of the outer region of the turbulent boundary layer. The advantage of using this set of parameters to describe the thickness ratio is that we already know that $\delta_1$ is a similarity parameter [11]. Hence, it is only necessary to prove that $\alpha_1$ is a similarity parameter in order to show that $\delta_d$ is also a similarity parameter.

## 3. Similarity Parameters of the 2-D Wall-bounded Flow

In the last section, the thickness characterization parameters for the inner and outer region velocity profile of the wall-bounded turbulent boundary layer were introduced. It turns out this moment integral based idea of the boundary layer can be adapted to study similarity issues [11]. For this case, we ask what can be determined about the scaled velocity profile in terms of integral moments of the **similarity defining equation**.

For 2-D wall-bounded flows, velocity profile similarity is defined to be where two velocity profiles taken at different stations along the flow differ only by simple scaling parameters in $y$ and $u(x,y)$. According to Schlichting [17], a velocity profile at position $x_i$ is similar to the velocity profile at $x_j$ if

$$\frac{u(x_i, y/\delta_s(x_i))}{u_s(x_i)} = \frac{u(x_j, y/\delta_s(x_j))}{u_s(x_j)} \quad \text{for all } y, \tag{8}$$

where the length scaling parameter is $\delta_s(x)$, and the velocity scaling parameter is $u_s(x)$. Although the intent of Schlichting's Eq. 8 expression is clear, it is based on redefining the velocity $u(x_i, y)$. In fact, the Schlichting's definition is a condensed way of expressing velocity profile similarity as

$$\frac{\overline{u}(x_i, y_i)}{u_s(x_i)} = \frac{\overline{u}(x_j, y_j)}{u_s(x_j)} \quad \text{where} \quad y_i = y_j, \quad \overline{u}(x_k, y_k) = u(x_k, y), \quad \text{and} \quad y_k = \frac{y}{\delta_s(x_k)} \quad \text{for all } y. \tag{9}$$

This definition emphasizes the fact that we are comparing the scaled velocity values at equivalent $y_i$-values and not equivalent $y$-values. For the work herein, we will use Eq. 9 as the operational definition for velocity profile similarity.

Starting with Eq. 9, Weyburne [11] developed a new way to study similarity. The new approach is based on taking integral moments of both sides of Eq. 9. By taking various integral moments, it is possible to find new properties of the scaled profiles which must be true if similarity is present in a set of velocity profiles. Weyburne [11] used this method, for example, to show that for any 2-D flow displaying similarity, the length scaling parameter $\delta_s(x)$ must be proportional to the velocity displacement thickness, $\delta_1(x)$, and the velocity scaling parameter $u_s(x)$ must be proportional to the velocity $u_e(x)$, the velocity at the boundary layer edge. In the next section, will use this method to show that if similarity is present then the thickness parameters $\delta_v$ and $\delta_d$ must be similarity scaling parameters.

### 3.1 Similarity parameters of the inner region

One method of verifying that the inner region boundary layer thickness, $\delta_v$, is a similarity parameter is to show that $\mu_1$ and $\sigma_v$ in Eq. 6 are similarity parameters. To do that we will work with the second derivative of Eq. 9. If the velocity profiles show similarity (Eq. 9 being true) then it follows that if we take the second derivative of both sides, then both sides must still be equal. If multiply the second derivatives of both sides by $y_i^n$, then both sides must still be equal. Now if we integrate both sides with respect to $y_i$ then they remain equal and we end up with moment-like integrals. Using these moment like integrals, one can obtain useful information about similarity of the velocity profile.

We will begin by considering the question of whether $\mu_1$ is a similarity parameter. Assume Eq. 9 is true for a set of velocity profiles. This means that $\delta_s$ and $u_s$ are similarity scaling parameters. If this is true, then the integral of the second derivative of both sides must equal. The integral of the second derivative is given by

$$a(x_i) = \int_0^h dy_i \frac{d^2\{\bar{u}(x_i,y_i)/u_s(x_i)\}}{dy_i^2} \tag{10}$$

$$a(x_i) = \frac{1}{u_s(x_i)} \int_0^h d\left\{\frac{y}{\delta_s(x_i)}\right\} \frac{d\bar{u}(x_i,y_i)}{d\left\{\frac{y}{\delta_s(x_i)}\right\}^2} = \frac{\delta_s(x_i)}{u_s(x_i)} \int_0^{h_i} dy \frac{d^2 u(x_i,y)}{dy^2}$$

$$a(x_i) = \frac{\delta_s(x_i)}{u_s(x_i)}\left[\frac{du(x_i,y)}{dy}\right]_{y=h_i,0}$$

$$a(x_i) = -\frac{\delta_s(x_i)}{u_s(x_i)}\frac{u_\tau^2(x_i)}{v},$$

where $h$ is located deep in the free stream above the wall and $h = h_i/\delta_s(x_i) = h_j/\delta_s(x_j)$. Similarity requires that $a(x_i) = a(x_j)$. The importance of this equation is that if similarity is present in a set of velocity profiles for any 2-D wall bounded flow, then the ratio of the thickness scaling parameter to the velocity scaling parameter must be inversely proportional to the friction velocity squared divided by the kinematic viscosity.

Furthermore, using Eq. 2, this means that

$$a(x_i) = a(x_j) \tag{11}$$

$$-\frac{\delta_s(x_i)}{u_s(x_i)}\frac{u_\tau^2(x_i)}{v} = -\frac{\delta_s(x_j)}{u_s(x_j)}\frac{u_\tau^2(x_j)}{v}$$

$$\frac{\delta_s(x_i)}{u_s(x_i)}\frac{u_e(x_i)}{\mu_1(x_i)} = \frac{\delta_s(x_j)}{u_s(x_j)}\frac{u_e(x_j)}{\mu_1(x_j)}$$

$$\frac{\delta_s(x_i)}{\mu_1(x_i)} = \frac{\delta_s(x_j)}{\mu_1(x_j)}.$$

Here we have used the fact that if similarity is present then $u_e/u_s$ is a constant [11]. The importance of this equation is that if similarity is present in a set of velocity profiles for any 2-D wall bounded flow, then the thickness scaling parameters must be proportional to the viscous mean location $\mu_1$.

Now we show that $\sigma_v$ is also a similarity parameter. Because $\delta_1$ is a similarity parameter [11] and we just showed that $\mu_1$ is a similarity parameter, therefore $\sigma_v$ must be a similarity parameter since $\sigma_v = \sqrt{-\mu_1^2 + 2\mu_1\delta_1}$. It follows from Eq. 3 that because $\mu_1$ and $\sigma_v$ are similarity parameters, $\delta_v$ must also be a similarity parameter.

## 3.2 Similarity parameters of the outer region

It is also possible to describe the outer boundary layer region with either the first derivative or the velocity profile based moments. The first derivative based moments have an advantage for this application because we already know that $\delta_1$ is a similarity parameter [11]. This means that it is only necessary to show that $\sigma_s$ is a similarity parameter in order to show $\delta_d$ is also a similarity parameter.

Starting with Eq. 9; if we take the derivative of both sides with respect to $y_i$, multiply both sides by $y_i^2$, and then integrate, both sides must still be equal. Defining $b(x_i)$ as the result of these operations, we have

$$b(x_i) = \int_0^h dy_i \; y_i^2 \frac{d\{\bar{u}(x_i,y_i)/u_s(x_i)\}}{dy_i} \tag{12}$$

$$b(x_i) = \frac{1}{u_s(x_i)}\int_0^h d\left\{\frac{y}{\delta_s(x_i)}\right\} \left\{\frac{y}{\delta_s(x_i)}\right\}^2 \frac{d\bar{u}(x_i,y_i)}{d\left\{\frac{y}{\delta_s(x_i)}\right\}} = \frac{1}{u_s(x_i)\delta_s^2(x_i)}\int_0^{h_i} dy \; y^2 \frac{du(x_i,y)}{dy}$$

$$b(x_i) = \frac{1}{u_s(x_i)\delta_s^2(x_i)}\left[y^2 u(x_i,y)\right]_{y=h_i,0} - \frac{2}{u_s(x_i)\delta_s^2(x_i)}\int_0^{h_i} dy \; y u(x_i,y)$$

$$b(x_i) = \frac{h_i^2 u_e(x_i)}{u_s(x_i)\delta_s^2(x_i)} - \frac{2}{u_s(x_i)\delta_s^2(x_i)}\int_0^{h_i} dy \; y u(x_i,y)$$

$$b(x_i) = \frac{2u_e(x_i)\alpha_1(x_i)}{u_s(x_i)\delta_s^2(x_i)} \; .$$

Similarity requires $b(x_i) = b(x_j)$ and we know that if similarity is present, then $u_e/u_s$ is a constant, so we have

$$\frac{\alpha_1(x_i)}{\delta_s^2(x_i)} = \frac{\alpha_1(x_j)}{\delta_s^2(x_j)} \; . \tag{13}$$

The importance of this equation is that if similarity is present in a set of velocity profiles for any 2-D wall bounded flow, then the parameter $\alpha_1$ must be a similarity parameter.

We know that $\delta_1$ is a similarity parameter [10] and we just showed that $\alpha_1$ is a similarity parameter, hence $\sigma_s$ must be a similarity parameter because $\sigma_s(x) = \sqrt{-\delta_1^2 + 2\alpha_1}$. Since $\delta_1$ and $\sigma_s$ are similarity parameters, it therefore follows from Eq. 6 that $\delta_d$ must also be a similarity parameter.

## 4. Discussion

It is generally accepted that the wall-bounded turbulent boundary layer velocity profiles do not show whole profile similarity, likely due to the viscosity effects in the inner region area [1]. As a result, it has been common to consider similarity issues for the inner and outer region separately [2]. The recent work of Weyburne [7-9] indicates that most experimental wall-bounded turbulent boundary layer datasets do NOT show similarity in the outer region. Hence, there is a question as to whether the outer region of the turbulent boundary layer can show similarity and whether there are any simple flow criteria that predict the flow conditions necessary for this to occur. In a work to be published [18], we show that there are cases of outer region wall-bounded velocity profile similarity and that the thickness ratio $\delta_d/\delta_v$ is useful for identifying likely experimental datasets.

An important aspect of this equal area similarity method, which was used to derive the new thickness ratio $\delta_d/\delta_v$, is that there are no approximations or assumptions. In this method, there is nothing equivalent to the **stream function assumption** regarding separable *x* and *y* functionals for the velocity required in the flow governing equation approach to similarity. In contrast, the equal area similarity method is based strictly on the mathematics of the **equation that defines similarity**. If we accept Eq. 9 as the definition of similarity, and accept that the boundary layer limits are $u(x_i,0)=0$ and $u(x_i,y)=u_e(x_i)$ for $y>\delta_d$, then the results obtained above are incontrovertible. Although not presented as such, all of the results obtained above can be put in the form of mathematical proofs.

## 5. Conclusions

If similarity is present in a set of wall-bounded turbulent boundary layer velocity profiles then it was demonstrated that the definitions for the inner and outer regions thickness, $\delta_v$ and $\delta_d$, are similarity scaling parameters. An important aspect of these parameters is that they are experimentally accessible.

### Acknowledgement

The author acknowledges the support of the Air Force Research Laboratory and Gernot Pomrenke at AFOSR.### References

[1] A. Townsend, *The Structure of Turbulent Shear Flows*, 2nd ed., Cambridge Univ. Press, Cambridge, U.K., 1976.

[2] W. George and L. Castillo, "Zero-Pressure Gradient Turbulent Boundary Layer," Applied Mechanics Reviews **50**, 689(1997).

[3] L. Castillo and W. George, "Similarity Analysis for Turbulent Boundary Layer with Pressure Gradient: Outer Flow," AIAA J. **39**, 41 (2001).8

**Appendix A**

Our contention that the Rotta [10] constraint, $u_e/u_\tau$, has never been theoretically proven has been contested in a recent paper by Jones, Nickels, and Marusic [19]. They published a paper in which they purport to offer a theoretical proof that the friction velocity $u_\tau(x)$ is a valid similarity scaling parameter for the turbulent boundary layer outer region. Since $u_e(x)$ is already a known similarity scaling parameter [1,3,10,11], then this would seem to disprove our contention. Their proof is based on the asymptotic infinite Reynolds number approach developed by George and Castillo [2]. Jones, Nickels, and Marusic adopt this approach in order to show that "… the classical scaling using the friction velocity also leads to a valid similarity solution for the outer flow in this limit." In reviewing this paper we found a critical flaw which invalidates their proof. Early on the paper assumes that Coles [20] Law of the Wall, Law of the Wake (in the form of their Eq. 3.18) is a valid equation as if it is based on a proven theory. A central defining feature of the Law of the Wall, Law of the Wake is that the velocity scaling parameter is the friction velocity $u_\tau(x)$. The problem is that the Law of the Wall, Law of the Wake has never been proven theoretically to apply to the turbulent boundary layer. In fact, that is exactly what Jones, Nickels, and Marusic are trying to prove. Hence, Jones, Nickels, and Marusic are assuming $u_\tau(x)$ is a valid similarity scaling parameter in order to prove that $u_\tau(x)$ is a valid similarity scaling parameter. This circular logic problem surfaces later in their paper when dealing with the infinite Reynolds number limit. According to their paper, in the infinite Reynolds number limit, their $a_2$ factor (=$1/C_1$) must be proportional to their $a_5$ factor (=1). This necessarily requires that $C_1$, their Eq. 3.22, must be a constant. Their $C_1$ is given by

$$C_1 = \int_0^1 d\eta \frac{U_1 - u}{u_\tau} \quad , \tag{A.1}$$

(where their $U_1$ is equal to $u_e$ as used above). Jones, Nickels, and Marusic assert that $C_1$ is a universal constant. They offer no proof for this assertion. In fact, $C_1$ only a constant at infinite Reynolds number if $u_\tau(x)$ is a valid similarity scaling parameter which, once again, is what they are trying to prove. Therefore, Jones, Nickels, and Marusic have used a circular logic argument in order to "prove" that the friction velocity $u_\tau(x)$ must be a valid similarity scaling parameter. A circular logic approach does not prove anything. Hence, our contention that the Rotta constraint has not been proven still applies.